\documentclass[11pt]{article}
\usepackage{amsmath, graphicx}

     \renewcommand{\baselinestretch}{1.3}
  \renewcommand{\arraystretch}{1.1}
\begin{document}

 \title{On Two Conversion Methods of Decimal-to-Binary}

 \author{Zhengjun Cao \\
  {\small Department of Mathematics, Shanghai University, Shanghai,
  China.}\\
  \textsf{  \textsf{caozhj@shu.edu.cn}}
   }

 \date{}\maketitle

\begin{abstract} Decimal-to-binary conversion  is  important to modern binary
 computers.
The classical  method to solve this problem is based on division operation. In this paper, we investigate a decimal-to-binary conversion
method based on addition operation. The method is very easily implemented by software.  The cost analysis shows that the latter is
more preferable than the classical  method.
Thus the current Input/Output translation hardware to convert between the internal digit pairs and the external standard BCD codes can be reasonably removed.

\end{abstract}

\section{Introduction}

 \begin{quote}
  \emph{IF OUR ANCESTORS had invented arithmetic by counting with their two fists or their eight fingers, instead of their ten ``digits", we would never have to
 worry about writing binary-decimal conversion routines.}  --- Donald E. Knuth
  \end{quote}

We know  a digit $a$ $\in \{0, 1, \cdots, 9\}$  received from a numeric key on a keyboard (see Picture 1) is represented as a string of 4 bits. The mechanism is usually referred to as Binary Coded Decimal (BCD for short). For instance, the digit 9 is represented  as
$\fbox{1001}$. When we input the integer $79$ using numeric keys `7' and `9', it  is first represented  as two strings
$\fbox{0111}\  \fbox{1001}$. We here omit the other leftmost bits in each byte which represent numeric type or sign.  The two strings  then will be converted into the resulting  binary representation $\fbox{1001111}$.

In 1988. B. Shirazi  et al \cite{SY88} investigated the problem of  VLSI designs for redundant binary-coded decimal addition.
In 2006, A. Kaivani  et al \cite{KA06} discussed the conversion  methods of Decimal-to-Binary. In the same year,   H. Thapliyal et al \cite{TA06} focused on the integration of
 BCD adder with CMOS.

In this paper, we investigate  another
method  to convert decimal input data into binary form. It takes full use of  Binary Coded Decimal mechanism to completely eliminate operations of comparison. It is directly based on addition operation and bit operation---shift.  The method is very easily implemented by software.
The cost analysis shows that this method is more preferable than  the classical  method.

\section{Conversion  based on Division Table}

There are basically two ways to transform a decimal number into a binary number: comparison with descending powers of two and subtraction and short division by two with remainder.

The first method proceeds as follows.   Use a base 2 table from right to left.
 Given a decimal number for conversion, find the greatest power of 2 that can be included in it. Write 1 for the binary digit first on the left and then subtract that number from the decimal one.  Move to the next power of two, record the identifier with 0 (if the rest cannot contain the next power of two) or 1 (if it can contain it).
Continue making the same comparison and subtractions until there are  nothing left. List all 1 and 0 symbols orderly.

Another method to identify the binary counterpart of a decimal number is through division by two. Here are the steps that you will have to follow:

1. You will have to divide your number by two. But first, to succeed into visualizing your work, write down 2 followed by a parenthesis and then the number you want to convert.

2. The quotient then must be written under the long division symbol, and, next to it the remainder from the operation. (If the number divides to two perfectly, the remainder will be 0, if it does not, it will be 1).

3. You will continue to repeat the operation, always writing down the remainders.

4. In the end, you will have a column of 1 and 0 which you will have to read from the bottom to the top. This will be your binary counterpart for the decimal number you wanted to convert.

The process of converting decimal input data into binary form  is generally accomplished by the following method which  is based on  division  operation \cite{Knuth2}.
  Given an integer $u$, we can obtain its binary representation $(\cdots U_2U_1U_0)_2$  as follows:
 $$U_0= u\, \mbox{mod}\, 2, \hspace*{4mm} U_1=\lfloor u/2 \rfloor\, \mbox{mod}\, 2, \hspace*{4mm} U_2=\lfloor \lfloor u/2 \rfloor/2\rfloor\, \mbox{mod}\, 2, \hspace*{4mm}\ldots,$$
 stopping when $\lfloor\ldots \lfloor \lfloor u/2 \rfloor/2\rfloor\ldots/2\rfloor=0 $.

Unlike the general division operation, the conversion of decimal-to-binary only requires a special division because an input number is represented by  Binary Coded Decimal mechanism in common PC.
 Concretely speaking, the conversion of $\fbox{0111}\  \fbox{1001}$ into $\fbox{1001111}$ could be accomplished as follows.
 \begin{itemize}

 \item[1.] Set up the following division table

    \begin{center}{\small
  \begin{tabular}{|c|c|l|l|}
   \hline
decimal  integer & borrowed sign +  BCD code & quotient & residue  \\  \hline
10& $\fbox{1}+\fbox{0000}$ & $\fbox{0101}$ & $\fbox{0000}$ \\
11& $\fbox{1}+\fbox{0001}$ & $\fbox{0101}$ & $\fbox{0001}$  \\
 12&  $\fbox{1}+\fbox{0010}$ & $\fbox{0110}$ & $\fbox{0000} $ \\
 13&  $\fbox{1}+\fbox{0011}$ & $\fbox{0110}$ & $\fbox{0001} $ \\
14&  $\fbox{1}+\fbox{0100}$ & $\fbox{0111}$ & $\fbox{0000}$  \\
 15&  $\fbox{1}+\fbox{0101}$ & $\fbox{0111}$ & $\fbox{0001}$  \\
16&  $\fbox{1}+\fbox{0110}$ & $\fbox{1000}$ & $\fbox{0000}$  \\
 17&  $\fbox{1}+\fbox{0111}$ & $\fbox{1000}$ & $\fbox{0001}$  \\
 18& $\fbox{1}+\fbox{1000}$ & $\fbox{1001}$ & $\fbox{0000}$  \\
 19& $\fbox{1}+\fbox{1001}$ & $\fbox{1001}$ & $\fbox{0001}$  \\   \hline
 \end{tabular} }\vspace*{3mm}

 Table 1:  Division Table
 \end{center}

 \item[2.] Upon  the second digit  is input,  take the rightmost bit 1 of the first digit $0111$
  as the first borrowed sign  and  make a right shift  to  generate the corresponding
 quotient.
 $$ \fbox{0111}\stackrel{\mbox{right shit}}{------- \longrightarrow} \ \mbox{quotient}\ \fbox{0011}, \ \mbox{borrowded sign}\ 1.   $$
Combing the first borrowed sign 1 with the second digit $1001 $,  look up the above Division Table to
get the corresponding quotient  and residue. Then  generate the second borrowed sign  by extracting the rightmost bit of the residue.
   \begin{equation*}
  \fbox{1}+ \fbox{1001}\stackrel{\mbox{look up Division Table}}{--------- \longrightarrow}
  \left\{
 \begin{aligned}
 & \mbox{quotient}\ \fbox{1001}, \\
  & \mbox{residue}\ 0001\longrightarrow \mbox{borrowed sign}\  1
     \end{aligned}
     \right. \end{equation*}
  Take the second borrowed sign  as the rightmost bit of the binary representation of the input if there is not a third digit.

    \item[3.] Apply the same operations to the intermediate quotient 39 of 79 divided by 2 which is directly composed by $\fbox{0011}$ and  $ \fbox{1001} $.
    By the same procedure, we will obtain the whole bit string of 79.  The process is depicted by the following Graph 1.
 \end{itemize}

 \emph{Cost analysis}.
  There are three type operations in the conversion method  based on Division Table. The first  is right shift of 1 bit.
 The second is to extract  the rightmost bit of a 4-bit string to obtain the corresponding borrowed sign.
 The third is to look up the Division Table and copy the corresponding quotient  if the borrowed sign  is `1'.

  If there are $n$ digits in an input, to obtain the rightmost bit of the resulting string
   the process requires about $n/2$ shift instructions and $n/2$ operations of extracting borrowed sign.
   Since the resulting string  has about $3n$ bits, it requires
   $\mathcal{O}(n^2)$ shift instructions, $\mathcal{O}(n^2)$ operations of looking up the Division Table  and $\mathcal{O}(n^2)$ operations of extracting borrowed sign.

 \vspace*{-60mm}
\hspace*{-8mm}\begin{minipage}{\linewidth}

\includegraphics[angle=0,height=20cm,width=18cm]{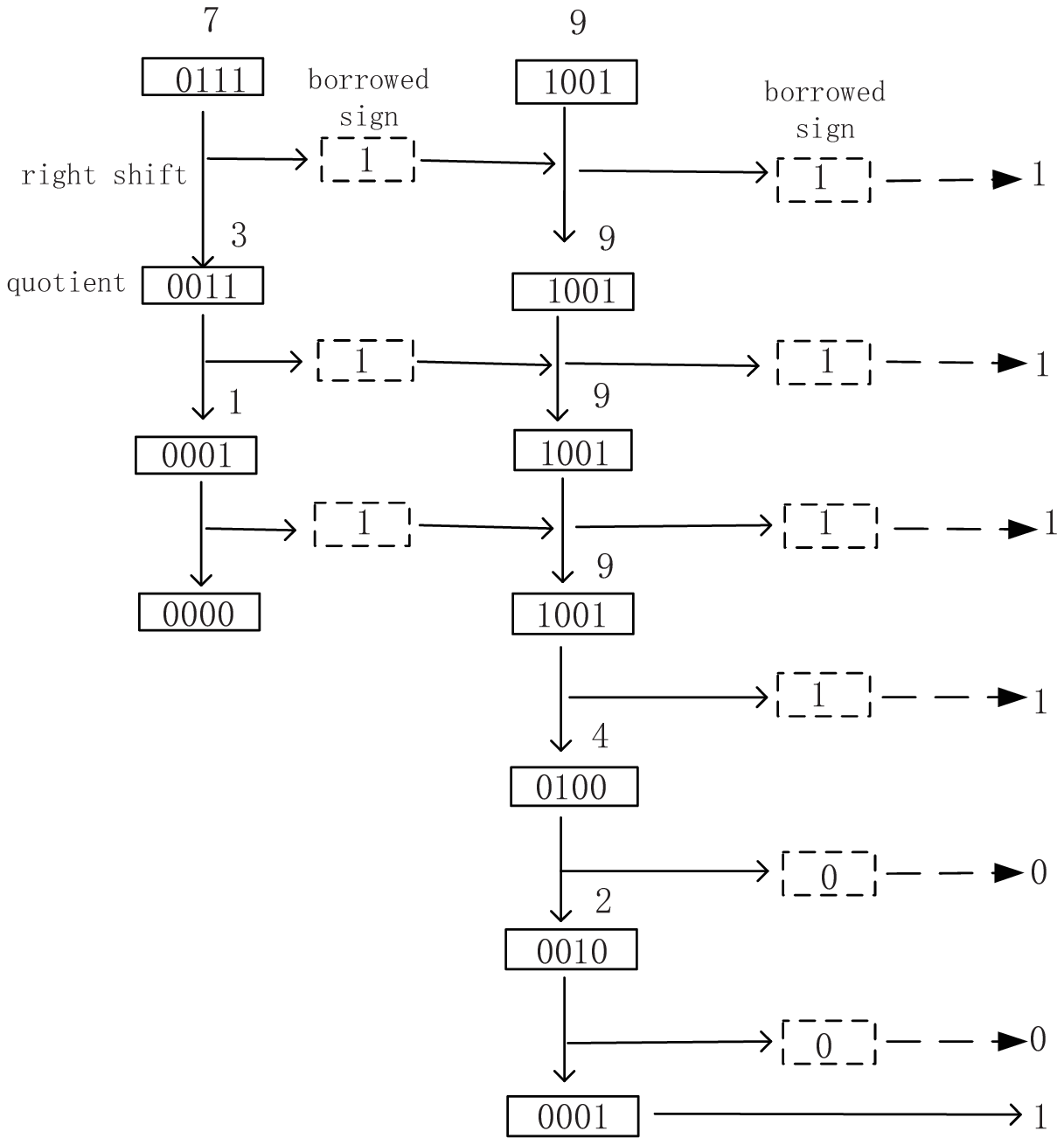}

 \end{minipage} \vspace*{-40mm}

 \centerline {Graph 1: Conversion of decimal-to-binary based on Division Table}

\section{Conversion  based on addition operation}

Although the general conversion method can be integrated into some hardware, the quantity of bit operations is impressive.
We now investigate another method of conversion of decimal-to-binary which is based on addition operation. The basic idea behind this method is that an integer can be written as
$$X_n\cdots X_2 X_1 =(\cdots ((X_n\times 10 +X_{n-1})\times 10+X_{n-2})\times 10+ \cdots )\times 10+X_1, \eqno(1) $$
where $X_i\in \{0, \cdots, 9\}, i=1, \cdots, n. $ Notice that  $$a\times 10=a\times 2^3+ a\times 2. $$

For illustration, if the digits $a=\fbox{0111},\  b=\fbox{1001}$ are received from two numeric keys, then a computer proceeds as follows.
\begin{itemize}
\item[1.]
Discard the leftmost bit 0 of the first string $\fbox{0111}$.   Then concatenate  ``0" and ``000"  to it, separately. It will generate two new strings
$$a_1=1110, \qquad  a_2=111000.  $$

\item[2.] Compute the addition $a'=a_1+a_2=1000110$. (Note that $a'=a\times 10$.)

\item[3.] Compute the resulting binary representation $s=a'+b=1000110+1001=1001111$.
\end{itemize}

 \emph{Cost analysis}.
   In the above procedure it requires one type of bit operation, shift. It also requires one type of  arithmetic operation, addition.
      To obtain $a'$ the process requires  2 shift instructions, 7  operations for computing
      $b_0+b_1+\epsilon $ where $b_0, b_1, \epsilon \in \{0, 1 \}$, $\epsilon$ is a carry bit.
      On average,  if there are $n$ digits in an input  the process  requires
       $\mathcal{O}(n)$ shift instructions, $\mathcal{O}(n^2)$ operations for computing
      $b_0+b_1+\epsilon $.  See the following table 2 for the cost comparison between the Method 1 and Method 2.

      \begin{center}{\small
      \begin{tabular}{|l|l|}
        \hline
          Method 1       &   Method 2 \\ \hline
        $\mathcal{O}(n^2) $ shift instructions &  $\mathcal{O}(n) $ shift instructions \\
       $\mathcal{O}(n^2) $ operations for looking up Division Table  &  $\mathcal{O}(n^2) $ operations for computing
      $b_0+b_1+\epsilon $  \\
       \qquad\quad    & \qquad  where $b_0, b_1, \epsilon \in \{0, 1 \}$, \\
         $\mathcal{O}(n^2) $ operations for extracting  borrowed signs &  $\mathcal{O}(n^2) $ operations for determining  carry bit $\epsilon$    \\
        \hline
      \end{tabular} }

      \vspace*{3mm}
      Table 2: Comparison between Method 1 and Method 2
       \end{center}

      In theory, the cost of an operation for looking up Division Table is greater than that of  an operation for computing
      $b_0+b_1+\epsilon $ where $b_0, b_1, \epsilon \in \{0, 1 \}$. Thus,
       the conversion method based on addition operation is more efficient than the method based on Division Table.

 Method 2 has a special advantage over Method 1. We can use Method 2 \emph{in parallel} because Eq.(1) can be written as
\begin{eqnarray*}
X_n\cdots X_2 X_1 &=& (X_n\cdots X_{i+1})\times 10^{i}+ X_i\cdots X_1 \\
&=&
[(\cdots ((X_n\times 10 +X_{n-1})\times 10+X_{n-2})\times 10+ \cdots )\times 10+X_{i+1}]\times 10^i \\
&+& (\cdots ((X_i\times 10 +X_{i-1})\times 10+X_{i-2})\times 10+ \cdots )\times 10+X_{1}.
\end{eqnarray*}
Whereas, Method 1 can not be used in such way.

\section{Conclusion}

We investigate  two conversion methods of decimal-to-binary.
 The analysis shows that the conversion method based on addition operation is more preferable than the general method which is based directly on division operation.
Thus the current Input/Output translation hardware to convert between the internal digit pairs and the external standard BCD codes can be reasonably removed.

\end{document}